\documentstyle[aps,epsfig,amsmath,amssymb]{revtex}

\draft

\begin{document}

\title{Moment of Inertia and Quadrupole Response Function of a Trapped
Superfluid}

\author{F.~Zambelli and S.~Stringari}

\address{Dipartimento  di Fisica, Universit\`a di Trento,}
\address{and Istituto Nazionale per la Fisica della Materia, 
I-38050 Povo, Italy}
\date{\today}

\maketitle

\begin{abstract}
We derive an explicit relationship between the moment of inertia
and the quadrupole response function of an interacting gas confined 
in a harmonic trap. The relationship holds for both Bose and
Fermi systems and is well suited to reveal the effects of
irrotationality of the superfluid motion. 
Recent experimental results on the scissors mode are used to extract
the value of the moment of inertia of a trapped Bose gas and to point
out the deviations from the rigid value due to superfluidity. 
\end{abstract}

\pacs{PACS numbers: 03.65.-w, 05.30.Jp, 32.80.Pj, 67.40.-w}


Important evidences of superfluidity in Bose-Einstein condensed gases
have become available in a recent series of experiments.
These include the occurrence of quantized vortices \cite{Jila,Dalibard},
the existence of a critical velocity for the motion of external
impurities \cite{Raman}, and the rotational properties
associated with the scissors mode \cite{Marago}. In particular
this latter experiment has revealed the consequences \cite{David} of
the quenching of the moment of inertia with respect to the classical
rigid value. 
The reduction of the moment of inertia follows from 
irrotationality which makes it impossible, for a superfluid, to
respond to a transverse probe and to rotate in a rigid way.
The moment of inertia characterizes the response of the system to a
rotational field $-\Omega J_z$ and is defined by the ratio 
\begin{equation}
\Theta =\frac{\langle J_z\rangle}{\Omega}
\label{theta1}
\end{equation} 
between the angular momentum $\langle J_z\rangle$ induced by the
rotation (hereafter assumed to take place around the $z$-axis)
and the angular velocity $\Omega$. 
In superfluid helium the moment of inertia was determined long time ago
by rotating the bucket containing the liquid and by measuring the
angular momentum of the sample \cite{Hess}. 
This experiment pointed out the occurrence of significant deviations 
from the rigid value for temperatures below the $\lambda$-point. 
Relation (\ref{theta1}) naturally defines the moment of inertia in the 
linear regime. At higher angular velocities new phenomena, associated 
with the creation of quantized vortices, take place in superfluids 
\cite{Donnelly}. Like in superfluid helium, also in a cold trapped gas 
the moment of inertia is expected to deviate significantly from the 
rigid value. 
The temperature dependence of $\Theta$ for a trapped Bose gas below the 
critical temperature $T_c$ for Bose-Einstein condensation and the 
corresponding deviations from the rigid value were investigated in 
\cite{Giorgini}.
First estimates of the moment of inertia of a trapped Fermi gas below
the BCS transition have been also reported \cite{Schuck}.
In a trapped gas one can rotate the confining potential using
magnetic or optical methods \cite{Foot,Dalibard}. 
Actually this procedure has proven quite successful in producing quantized
vortices in Bose-Einstein condensed gases \cite{Dalibard}.
However, the direct measurement of the angular momentum induced by the
rotation of the trap is difficult in atomic gases, since
most of diagnostic techniques, based on optical imaging, provide
information on the density profiles, either {\em in situ} or after the 
expansion of the gas. 

In this work we propose a method to measure the moment
of inertia of a trapped gas based on the study of the collective
oscillations resulting from the coupling between the rotational 
and the quadrupole motion. 
Our analysis, which develops the ideas introduced in \cite{David}, is
based on the derivation of an explicit relationship
between the moment of inertia and the quadrupole response
function, and will be applied to the recent
experiments on the scissors mode of a trapped Bose gas \cite{Marago}.
The possibility of revealing the effects of superfluidity using this
method would be particularly appealing in the case of Fermi 
systems, where most of diagnostic techniques are insensitive to the
occurrence of the BCS transition.

The starting point of our analysis is the commutation rule ($\hslash =1$)
\begin{equation}
[H,J_z] = -i(x\nabla_y-y\nabla_x)V_{\rm ext}({\mathbf r})\;, 
\label{[HJ]general}
\end{equation}
where $J_z=\sum_{i=1}^Nx_ip_i^y-y_ip_i^x$ is the third component of
the angular momentum operator and
\begin{equation}
H = \sum_{i=1}^N\frac{p_i^2}{2M}+\sum_{i<j}V({\mathbf r}_i-{\mathbf
  r}_j)+V_{\rm ext}({\mathbf r})\;, 
\label{H}
\end{equation} 
is the Hamiltonian of the system, containing the external confining potential 
$V_{\rm ext}({\mathbf r})$. Result (\ref{[HJ]general}) is independent
of the choice of the two-body potential, and follows from the
rotational symmetry of the internal Hamiltonian $H-V_{\rm ext}$. 
In the following we will consider harmonic potentials
\begin{equation}
V_{\rm ext}({\mathbf r})=\sum_{i=1}^N
\frac{M}{2}(\omega_x^2x_i^2+\omega_y^2y_i^2+\omega_z^2z_i^2) 
\label{Vext}
\end{equation}
with an asymmetry $(\omega_x\neq\omega_y)$ in the rotating plane. 
In this case the commutation rule (\ref{[HJ]general}) takes the simple
form
\begin{equation}
[H,J_z] = -iM(\omega^2_x-\omega_y^2)Q\;, 
\label{[HJ]}
\end{equation}
where
\begin{equation}
Q=\sum_{i=1}^Nx_iy_i 
\label{Q}
\end{equation}
is the relevant quadrupole operator. 
Eqs. (\ref{[HJ]})-(\ref{Q}) explicitly reveal the coupling between the degrees
of freedom associated with the angular momentum and quadrupole
variables, induced by the deformation of the confining potential.

For a sample in thermal equilibrium at temperature $T$ the moment of
inertia is given, according to linear response theory, by the 
expression
\begin{equation}
\Theta =\frac{1}{\cal{Z}}\sum_{n,m}\big[e^{-\beta\omega_m}
-e^{-\beta\omega_n}\big]
\frac{|\langle m|J_z|n\rangle |^2}{\omega_n-\omega_m}\;,
\label{I}
\end{equation}
where $|n\rangle$ and $\omega_n$ are eigenstates and
eigenfrequencies of the Hamiltonian (\ref{H}), $\cal{Z}$ is the
canonical partition function and $e^{-\beta\omega_n}$ is the usual
Boltzmann factor, with $\beta=1/k_BT$. 

Because of the algebraic relationship (\ref{[HJ]}) the moment of inertia
can be also written in the form 
\begin{equation}
\Theta=-\frac{M^2}{\pi}(\omega^2_x-\omega_y^2)^2 
\int\!d\omega\,\frac{\chi_Q^{\prime\prime}(\omega)}{\omega^3} 
\label{theta2}
\end{equation}
where
\begin{equation}
\chi_Q^{\prime\prime}(\omega)= 
\frac{\pi}{\cal{Z}}\sum_{n,m}[e^{-\beta\omega_n}-e^{-\beta\omega_m}
] |\langle m|Q|n\rangle |^2\delta (\omega -\omega_{nm})\;,
\label{ImchiQ}
\end{equation}
is the imaginary part of quadrupole dynamic response function
\begin{equation}
\chi_Q(\omega)=\frac{2}{\cal{Z}}\sum_{n,m}e^{-\beta\omega_m}
\frac{|\langle m|Q|n\rangle|^2\omega_{nm}}
{(\omega+i\eta)^2-\omega_{nm}^2}\;,
\label{chiQ}  
\end{equation}
and $\omega_{nm}=\omega_n-\omega_m$. 
Here and in the following the frequency integrals are taken
between $-\infty$ and $+\infty$. 
It is useful to compare the moment of inertia (\ref{theta2}) with the
rigid value
\begin{equation}
\Theta_{\rm rig}=MN\langle x^2+y^2\rangle
\label{Irigid}
\end{equation}
given by classical theory.
One can easily show that the average (\ref{Irigid}) is proportional to
the double commutator of the Hamiltonian (\ref{H}) with the quadrupole
operator (\ref{Q}). In fact the following sum rule holds \cite{Lipparini}:
$\langle [Q,[H,Q]]\rangle=N\langle x^2+y^2\rangle /M$. This result is
the natural extension of the dipole Thomas-Reiche-Kuhn sum rule \cite{TRK}
to the quadrupole operator. 
On the other hand the double commutator is directly related to the
energy weighted moment of the imaginary part (\ref{ImchiQ}) of the
quadrupole response function, so that one finally obtains the exact 
relationship $N\langle x^2+y^2\rangle=-M\int\!
d\omega\chi_Q^{\prime\prime}(\omega)\omega /\pi$,
which permits to write the ratio between the moment of inertia 
(\ref{theta2}) and the rigid value (\ref{Irigid}) in the useful form
\begin{equation}
\frac{\Theta}{\Theta_{\rm rigid}}=(\omega_x^2-\omega_y^2)^2
\frac{\int\!d\omega \chi_Q^{\prime\prime}(\omega)/\omega^3}
{\int\!d\omega \chi_Q^{\prime\prime}(\omega)\omega}\;.
\label{ratio1}
\end{equation}
Eq. (\ref{ratio1}) represents a major result of the present paper.
It explicitly shows that the occurrence of superfluidity,
characterized by a reduction of the value of the moment of inertia
with respect to the rigid value, must show up 
also in the structure of the quadrupole excitations of the system,
characterized by the function $\chi_Q^{\prime\prime}(\omega )$.
Actually the physical consequence of superfluidity is rather clear.
The constraint of irrotationality suppresses the modes with low
frequency ($\omega\sim|\omega_x-\omega_y|$). As a consequence in a
perfect superfluid the integral (\ref{theta2}) approaches a constant
value when $\omega_x\to\omega_y$ with the consequent quenching of the value 
of $\Theta$, which behaves like $(\omega_x^2-\omega_y^2)^2$.
The suppression of the low frequency oscillations in the presence of
Bose-Einstein condensation has been experimentally pointed out in  
\cite{Marago}. Conversely, in a normal system the occurrence of low
frequency modes is crucial to bring the moment of inertia to the rigid value.

It is worth discussing the effects of superfluidity more 
explicitly. The equations characterizing the macroscopic behaviour
of a superfluid at zero temperature have the form of irrotational
hydrodynamics 
\begin{align}
&\frac{\partial \rho}{\partial t}+{\boldsymbol\nabla}
(\rho{\mathbf v})=0
\label{continuity}\\
&\frac{\partial{\mathbf v}}{\partial
  t}+{\boldsymbol\nabla}\left(\frac{v^2}{2}+
\frac{V_{\rm ext}({\mathbf r})}{M}+\frac{\mu_{\rm loc}(\rho)}{M}\right)=0
\label{veq}
\end{align} 
where $\mu_{\rm loc}(\rho)$ is the chemical potential of a uniform system
evaluated at density $\rho$, and ${\mathbf v}$ is the irrotational
superfluid velocity. These equations hold in particular for a cold dilute
Bose gas in the Thomas-Fermi limit \cite{rmp}.
In this case $\mu_{\rm loc}(\rho)=g\rho$ is linear in the density, the coupling
constant $g=4\pi a/m$ being fixed by the $s$-wave scattering length $a$.
In this limit the hydrodynamic equations (\ref{continuity}-\ref{veq})
are equivalent to the Gross-Pitaevskii equation
for the order parameter and can be easily solved to
obtain the collective frequencies of a trapped Bose-Einstein condensed
gas \cite{SS}. 
In Ref. \cite{PS} the same Eqs. (\ref{continuity}-\ref{veq}) have been 
used to calculate the corrections to the collective frequencies
of a trapped Bose gas, due to the inclusion of the Lee-Huang-Yang term
\cite{LHY} in the equation of state.
Applicability of Eqs. (\ref{continuity}-\ref{veq}) is not however 
restricted to Bose superfluids. They are in fact expected to hold also
in the case of Fermi superfluids. Of course in this case the equation
of state can be very different due Pauli effects. 
For example, in a very dilute Fermi gas the chemical potential is a 
purely kinetic effect: $\mu_{\rm loc}(\rho)=(3\pi^2\rho)^{2/3}/2m$ 
(we consider here a Fermi gas of two spin species with equal 
density). 
In this limit the hydrodynamic equations of superfluids have been 
investigated in \cite{Baranov}. In general the applicability of
Eqs. (\ref{continuity}-\ref{veq}) is subject to the following
constraints: ({\em i}) the non-superfluid component of the system
should be negligible and hence the temperature should be much smaller
than the transition temperature to the normal phase; ({\em ii}) the
sample should be large enough in order to ensure the validity of the
local density approximation for the chemical potential (Thomas-Fermi limit); 
({\em iii}) the length scale of the oscillations should be larger than 
the healing length. 

Starting from the equations of irrotational hydrodynamics
it is straightforward to calculate
the moment of inertia. To this purpose it is convenient to rewrite these
equations in the frame rotating with angular velocity $\Omega$.
One finds
\begin{align}
&\frac{\partial \rho}{\partial t}+{\boldsymbol\nabla}
\big(\rho({\mathbf v}-{\boldsymbol\Omega}\times{\mathbf r})\big)=0
\label{Rcontinuity}\\
&\frac{\partial{\mathbf v}}{\partial
  t}+{\boldsymbol\nabla}\cdot\left(\frac{v^2}{2}+\frac{V_{\rm
  ext}({\mathbf r})}{M} +\frac{\mu_{\rm loc}(\rho)}{M}-{\mathbf
  v}\cdot({\boldsymbol\Omega}\times{\mathbf r})\right)=0\;. 
\label{Rveq}
\end{align} 
The stationary solutions in the rotating frame have the simple
irrotational form 
\begin{equation}
{\mathbf v}=\Omega\frac{\langle x^2-y^2\rangle}{\langle
  x^2+y^2\rangle}{\boldsymbol\nabla}(xy)\;.
\label{virr}
\end{equation}
From Eq. (\ref{virr}) one can easily calculate the angular momentum and,
using definition (\ref{theta1}), one finds the irrotational result \cite{BM}:
\begin{equation}
\Theta=\left[\frac{\langle x^2-y^2\rangle}{\langle
    x^2+y^2\rangle}\right]^2\Theta_{\rm rigid}
\label{thetairr}
\end{equation} 
for the moment of inertia, showing that in a superfluid the value of
the moment of inertia is smaller than the rigid value \cite{supernota}.
Results (\ref{virr}-\ref{thetairr}) are not restricted to the
perturbative regime of small $\Omega$. They are simply the
consequence of the fact that the equilibrium density, fixed 
by the stationary solution of the Euler equation
(\ref{Rveq}), depends on the variables $x$ and $y$ through the quadratic
combination $\alpha_xx^2+\alpha_yy^2$ \cite{Ripoll}.
For small angular velocities the moment of inertia (\ref{thetairr}) reduces to
\begin{equation}
\Theta=
\left[\frac{\omega_x^2-\omega_y^2}{\omega_x^2+\omega_y^2}\right]^2
\Theta_{\rm rig}\;.
\label{thetaHD}
\end{equation} 

Another important feature emerging from the hydrodynamic Eqs. 
(\ref{continuity}-\ref{veq}) is that also the behaviour of the surface
oscillations, characterized by the condition 
${\boldsymbol \nabla}\cdot{\mathbf v}=0$, is independent of the 
equation of state \cite{nota2}. 
In particular, by adding a term of the form 
$\lambda xye^{i\omega t}+\hbox{c.c.}$ in the external potential one
finds that the quadrupole operator (\ref{Q}) excites, in
the linear limit, only one mode with frequency 
\begin{equation}
\omega_{\rm HD}=\sqrt{\omega_x^2+\omega_y^2}\;, 
\label{freqHD}
\end{equation}
so that the imaginary part of the quadrupole response function takes
the simple form 
\begin{equation}
\chi_Q^{\prime\prime}=
-\frac{N\pi}{2}\frac{\langle x^2+y^2\rangle}{M\omega_{\rm HD}}
\big[\delta(\omega-\omega_{\rm HD})-\delta(\omega+\omega_{\rm HD})\big]\;.
\label{chi2Q}
\end{equation} 
By inserting result (\ref{chi2Q}) into Eq. (\ref{ratio1}) one 
immediately recovers the irrotational value (\ref{thetaHD})
for the moment of inertia.

In the second part of the work we show that the deviations of the
moment of inertia from the rigid value can be measured through
a rather straightforward experiment which consists of a sudden
rotation of the confining trap by a small angle, followed by the imaging
of the oscillations of the gas. This experiment has been
recently carried out \cite{Marago} in a trapped Bose gas where
the effects of superfluidity predicted in \cite{David} for the
scissors mode have been directly observed.

The abrupt rotation of the harmonic trap is characterized by the
transformation of coordinates 
$x\to x\cos{\phi_0}+y\sin{\phi_0}$, $y\to -x\sin{\phi_0}+y\cos{\phi_0}$. 
For small values of $\phi_0$, satisfying the condition 
$\phi_0\ll(\omega_x^2-\omega_y^2)/(\omega_x^2+\omega_y^2)$, the
corresponding change in the harmonic potential (\ref{Vext}) produces a
time dependent perturbation of the form
\begin{equation}
H_{\rm pert}=\lambda Q\theta (t)\;,
\label{Hpert}
\end{equation}
where $\lambda=M(\omega_x^2-\omega_y^2)\phi_0$, $Q$ is the quadrupole
operator (\ref{Q}) and $\theta(t)$ is the usual step 
function.
Eq. (\ref{Hpert}) can be rewritten in its Fourier components as 
\begin{equation}
H_{\rm pert}=\lambda Q\frac{i}{4\pi}\int\!d\omega \frac{e^{-i\omega t}} 
{\omega +i\eta} + \hbox{H.c.}\;.
\label{Hpert2}
\end{equation}
According to  linear response theory the average value
$Q(t)\equiv\langle Q\rangle$ of the quadrupole operator evolves, 
for $t>0$, as
\begin{equation}
Q(t)=\lambda\frac{i}{4\pi}\int\!d\omega \frac{e^{-i\omega t}}
{\omega+ i\eta}\chi_Q(\omega )+\hbox{c.c.}\;,
\label{Qresponse}
\end{equation}
where $\chi_Q(\omega )$ is the quadrupole response (\ref{chiQ}). 
Integrating (\ref{Qresponse}) in the complex plane and using the analytic
properties of $\chi_Q(\omega)$, one  can rewrite Eq. (\ref{Qresponse})
in the form
\begin{equation}
Q(t)=\int\!d\omega Q(\omega)e^{i\omega t}-
\int\!d\omega Q(\omega)\;,
\label{exp}
\end{equation}
with 
\begin{equation}
Q(\omega )=
-\frac{\lambda}{\pi}\frac{\chi_Q^{\prime\prime}(\omega )}{\omega}\;.
\label{iden}
\end{equation}
Combining the above results with Eq. (\ref{ratio1}), one finally finds 
the useful relationship
\begin{equation}
\frac{\Theta}{\Theta_{\rm rigid}}=(\omega_x^2-\omega_y^2)^2
\frac{\int\!d\omega Q(\omega)/\omega^2}
{\int\!d\omega Q(\omega)\omega^2}\;,
\label{ratio2}
\end{equation}
which provides the sought link between the moment of inertia of the
system and the Fourier transform of the observable
quadrupole signal 
\begin{equation}
Q(t)=\int\!d{\mathbf r}\rho({\mathbf r},t)xy\;.
\label{signal}
\end{equation}
In actual experiments the quantity which is more easily measured is 
the angle of rotation $\phi (t)$ rather than the quadrupole moment
(\ref{signal}).
The angle $\phi (t)$ fixes the position of the 
symmetry axis of the gas with respect to the new symmetry axis of  
the confining trap.
During the rotation the density of the gas varies according to the 
law, valid for small angles,
\begin{equation}
\rho({\mathbf r},t)=\rho_0({\mathbf r})+(x\nabla_y
-y\nabla_x)\rho_0({\mathbf r})(\phi(t)-\phi_0)\;,
\label{dens}
\end{equation}
where $\rho_0({\mathbf r})$ is the density before the
sudden rotation of the trap.
Using Eqs. (\ref{signal}) and (\ref{dens}) one derives the useful relationship
\begin{equation}
Q(t)=N\langle x^2-y^2\rangle(\phi (t)-\phi_0)\; 
\label{ident}
\end{equation}
between the angle of rotation and the quadrupole moment.

We are now ready to use our formalism to extract the moment of inertia 
from the experiment of Ref. \cite{Marago}. 
At low temperature the angle $\phi(t)$ was measured
at different times after the sudden rotation of the confining potential.
The experimental results have confirmed the occurrence of a single
frequency in $\phi (t)$, as predicted by hydrodynamic theory:
\begin{equation}
\phi(t)=\phi_0\cos({\omega_{\rm HD}t})\;.
\label{signalc}
\end{equation}  
The observed frequency $\omega$ coincides, with excellent
accuracy, with the hydrodynamic value (\ref{freqHD}). Using the
relationship (\ref{ident}) and 
inserting the experimental values $\omega_x=2\pi (90\pm 0.2)$ 
sec$^{-1}$, $\omega_y=2\pi (249\pm 0.6)$ sec$^{-1}$ and 
$\omega =2\pi (265.6\pm 0.8)$ sec$^{-1}$ into Eq. (\ref{ratio2}), one 
obtains the value  $\Theta /\Theta_{\rm rigid}\simeq 0.6$,
providing an unambiguous proof of the superfluidity of the sample.

At higher temperature, well above $T_c$, the situation changes
drastically. In first approximation the effects of interactions can be
neglected and the time evolution of $\phi(t)$ can be described by the
non-interacting model \cite{David}
\begin{equation}
\phi(t)=\frac{1}{2}\phi_0[\cos{(\omega_+t)}+\cos{(\omega_-t)}]\;,
\label{signalcl}
\end{equation}
where
\begin{equation}
\omega_{\pm}=|\omega_x\pm\omega_y|
\label{freqsclass}
\end{equation}
are the two frequencies excited by the external perturbation.
The experimental results of \cite{Marago}, carried out at $T\sim 5T_c$,
well agree with these predictions. In fact the observed frequencies
$\omega_+=2\pi (338.5\pm 0.8)$ sec$^{-1}$ and 
$\omega_-=2\pi (159.1\pm 0.8)$ sec$^{-1}$ are very close to the
theoretical values (\ref{freqsclass}) and enter the measured signal 
$\phi (t)$ with practically equal weight.
These results, inserted into Eq. (\ref{ratio2}), give finally
$\Theta /\Theta_{\rm rigid}\simeq 1$, 
confirming the crucial role played by the mode with the low frequency
$\omega=|\omega_x-\omega_y|$ in providing the rigid response to the
rotational field.

A similar behaviour is expected to occur also in a Fermi gas. In fact
at temperatures much smaller that the BCS transition the response is
still given by the hydrodynamic predictions (\ref{freqHD}) and
(\ref{signalc}). On the other hand, in the normal phase of a very
dilute Fermi gas one can use the predictions
(\ref{signalcl})-(\ref{freqsclass}) of the non-interacting model.
 
The above procedure can be also used to investigate the effects of
superfluidity as a function of temperature. In particular, by
investigating the measured signal (\ref{signal})
and inserting its Fourier transform into Eq. (\ref{ratio2}), one can
determine the value of the moment of inertia at any value of $T$.
One should however take into account that, below the critical
temperature, the density profile of the oscillating system cannot be 
simply parametrized through a single angle of rotation, like in
Eq. (\ref{dens}), because the normal and the superfluid
components will respond in a different way to the rotation of the trap. 
As a consequence one should introduce two distinct angles
and the density of the gas will oscillate, in the linear limit, 
according to
\begin{equation}
\rho({\mathbf r},t)=\rho_0({\mathbf r})+(x\nabla_y
-y\nabla_x)\rho_S({\mathbf r})\phi_S(t)+
(x\nabla_y-y\nabla_x)\rho_N({\mathbf r})\phi_N(t)\;,
\label{dens2}
\end{equation}
where $\rho_0({\mathbf r})=\rho_S({\mathbf r})+\rho_N({\mathbf r})$ is
the sum of the superfluid and normal component at equilibrium.
The careful investigation of the time evolution of the density profile
would provide unique information on the shape of the superfluid and
normal components. In a dilute Bose gas these components are expected
to coincide, respectively, with the condensate and the thermal cloud,
except for temperature smaller than the chemical potential. The
situation is very different in Fermi gases, where the total density
profile is scarcely affected by the BCS transition \cite{Stoof}, and
the separation between the two components is much less trivial. 

In conclusion we have shown that the moment of inertia $\Theta$ of a
gas trapped by harmonic potential can be explicitly measured by
analyzing the quadrupole oscillations generated by the a sudden rotation
of the trap. An explicit evaluation of $\Theta$ has been provided using
the recent data on the scissors mode available for a trapped Bose gas
either at very low temperature and above the critical temperature.
In the first case the method explicitly reveals the quenching of the
moment of inertia due to superfluidity. The same method can be
naturally applied to a Fermi system in order to reveal the effects of
superfluidity associated with the BCS transition. Moreover,
with suitable modifications, it can be also extended to the case of 
non-harmonic trapping.

Stimulating discussions with D.~Gu\'ery-Odelin are acknowledged. We
are also grateful to O.~Marag\`o for providing us the data of the
Oxford experiment on the scissor mode.


\begin{references}
\bibitem{Jila}
M.R.~Matthews, B.P.~Anderson, P.C.~Haljan, D.S.~Hall, C.E.~Wieman,
and E.A.~Cornell, Phys. Rev. Lett. {\bf 83}, 2498 (1999).

\bibitem{Dalibard}
K.W.~Madison, F.~Chevy, W.~Wohlleben, and J.~Dalibard,
Phys. Rev. Lett. \textbf{84}, 806 (2000).

\bibitem{Raman}
C.~Raman, M.~K\"ohl, R. Onofrio, D.S.~Durfee, C.E.~Kuklewicz,
Z.~Hadzibabic, and W.~Ketterle, Phys. Rev. Lett. \textbf{83}, 2502 (1999).

\bibitem{Marago}
O.M.~Marag\`o, S.A.~Hopkins, J.~Arlt, E.~Hodby, G.~Hechenblaikner, and
C.J.~Foot, Phys. Rev. Lett. \textbf{84}, 2056 (2000).

\bibitem{David}
D.~Gu\'ery-Odelin and S.~Stringari, Phys. Rev. Lett. \textbf{83},
4452 (1999).

\bibitem{Hess}
G.B.~Hess and W.M.~Fairbank, Phys. Rev. Lett. \textbf{19}, 216
(1967).

\bibitem{Donnelly}
R.J.~Donnelly, {\em Quantized vortices in Helium II} (Cambridge
University Press, Cambridge, 1995). 

\bibitem{Giorgini}
S.~Stringari, Phys. Rev. Lett. \textbf{76}, 1405 (1996);
S.~Giorgini, L.P.~Pitaevskii, and S.~Stringari, J. Low
Temp. Phys. \textbf{109}, 309 (1997).

\bibitem{Schuck}
M.~Farine, P.~Schuck, and X.~Vinas, e-print cond-mat/9901241, accepted
for publication in Phys. Rev. A. 

\bibitem{Foot}
J.~Arlt, O.~Marag\`o, E.~Hodby, S.A.~Hopkins, G.~Hechenblaikner,
S.~Webster, and C.J.~Foot, to be published J. Phys. B Jan 2000.

\bibitem{Lipparini}
E.~Lipparini and S.~Stringari, Phys. Rep. \textbf{175}, 103-261 (1989). 

\bibitem{TRK}
N.~Kuhn, Z. Phys. \textbf{33}, 408 (1923);
L.~Ladenburg and F.~Reiche, Naturwis. \textbf{11}, 873 (1923);
L.H.~Thomas, Naturwis. \textbf{13}, 627 (1925).

\bibitem{rmp} F.~Dalfovo, S. Giorgini, L. Pitaevskii and S. Stringari,
Rev. Mod. Phys. \textbf{71}, 463 (1999).

\bibitem{SS}
S.~Stringari, Phys. Rev. Lett. \textbf{77}, 2360 (1996).

\bibitem{PS}
L.~Pitaevskii and S.~Stringari, Phys. Rev. Lett. \textbf{81}, 4541 (1998).

\bibitem{LHY}
T.D.~Lee and C.N. Yang, Phys. Rev. \textbf{105}, 1119 (1957);
T.D.~Lee, K.W.~Huang, and C.N.~Yang, Phys. Rev. \textbf{106}, 1135 (1957).

\bibitem{Baranov}
M.A.~Baranov and D.S.~Petrov, e-print cond-mat/9901108.
The same equations have been also used
by M.~Amoruso {\em et al.},  Eur. Phys. J. D \textbf{7}, 441 (1999),
to investigate the collisional regime of a normal Fermi gas.

\bibitem{BM}
A.~Bohr and B.~Mottelson, {\em Nuclear Structure} (Benjamin, New York,
1975), Vol. 2.

\bibitem{Ripoll}
J.J.~Garc{\'\i}a-Ripoll and V.M.~P\'erez-Garc{\'\i}a, e-print
cond-mat/0003451 (2000).

\bibitem{supernota}
The irrotational value of the moment of inertia coincides with the
rigid one in the case of a very deformed sample ($\langle
x^2\rangle\gg\langle y^2\rangle$).

\bibitem{nota2}
In the presence of a deformed trap the surface excitations may be
coupled with the compressional modes and the corresponding dispersion
will depend on the equation of state. It is easy to show that the
surface modes are decoupled only for velocity fields satisfying the conditions 
${\boldsymbol\nabla}\cdot{\mathbf v}=0$ and 
$M\omega^2v_k=\nabla_k({\boldsymbol\nabla}V_{\rm ext}({\mathbf r})
\cdot{\mathbf v})$. For triaxially
deformed traps the above equations admit only a few solutions. These
include the quadrupole mode discussed in the text (with the natural
generalization to the $yz$ and $zx$ operators) as well as the
octupole mode ${\mathbf v}={\boldsymbol\nabla}(xyz)$ with 
$\omega^2=\omega_x^2+\omega_y^2+\omega_z^2$, in addition, 
of course, to the dipole modes along the three axes.

\bibitem{Stoof}
M.~Houbiers, R.~Ferwerda, H.T.C.~Stoof, W.I.~McAlexander,
C.A.~Sackett, R.G.~Hulet, Phys.\ Rev.\ A \textbf{56}, 4864 (1997). 

\end{references}
\end{document}